\documentclass[a4paper,11pt]{article}
\usepackage{graphicx}
\begin{document}
\bibliographystyle{plain}

\begin{flushleft}
{\Large
Neutrinos and (Anti)neutrinos from Supernovae 
and from the Earth in the Borexino detector
}
\vspace{1cm}
\\
Lino~Miramonti$^{1}$
\\
1) Physics Depart. Milano University and Istituto Nazionale di Fisica Nucleare  sez. Milano \\
\end{flushleft}
\vspace{0.5cm}
\begin{abstract}
The main goal of the Borexino detector, in its final phase of construction in the Gran Sasso underground laboratory, is the direct observation and measurement of the low energy component of neutrinos coming from the Sun. The unique low energy sensitivity and ultra-low background bring new capabilities to attack problems in neutrino physiscs other than solar ones. Investigation about the study of Supernoavae neutrinos and neutrino coming from the Earth (Geoneutrinos) are here resumed. This work is based on the works in references \cite{ArticoloCadonati} and \cite{calaprice}.
\end{abstract}

\section{Introduction}
The main goal of the Borexino detector is the direct observation and measurement, 
in real time, of the low energy component ($\leq$ 1 MeV) of  the neutrinos coming from the Sun; 
in particular the neutrino monochromatic line at 862 keV due to the electron capture of $^7Be$.

The Borexino design has been described elsewhere (see for instance \cite{STBX}). It is a large 
unsegmented calorimeter featuring 300 tons of liquid scintillator, contained in a 8.5 meter nylon 
vessel (125 $\mu$m thick), viewed by 2200 photomultiplier tubes.

The study of this low energy component is possible thank to the definition of a fiducial volume,  
featuring 100 tons of liquid scintillator, in order to maximize the signal to noise ratio mainly due to 
the photomultiplier tubes (i.e. esternal background).
The radiopurity level reach a value of the order of $10^{-16} g/g$  (U and Th equivalent) and 
$10^{-14} g/g$ in K (i.e. internal background).

The graded shield of progressively lower radioactivity material approaching the detector's core, 
the achievement of high radiopurity level and the definition of the fiducial volume allows a measurement 
as low as 0.1 events per day per ton via the $ \nu + e^- \rightarrow \nu + e^- $ electroweak scattering reaction.

The Borexino detector can be applied to a wide range of open questions in particle physics, astrophysics and geophysics.

\section{Antineutrino detection}
The best method to detect electron antineutrinos is the classic Cowan Reines reaction of capture by a proton in a liquid scintillator: $ \bar{\nu}_{e} + p \rightarrow n + e^+ $.

This reaction provides a coincidence tag that severely suppresses background against $ \bar{\nu}_{e} $ signals as 
small as events per year in a kiloton-scale liquid scintillator detector. The electron antineutrino tag is made possible
by a delayed coincidence of $e^{+}$ and by a 2.2 MeV $\gamma$-ray emitted by capture of neutron on a proton after a delay 
of $\sim$ 200 $\mu s$. The tag suppresses background by about 2 order of magnitude, hence, the entire scintillator mass of 
300 tons may be employed.

Between sources of correlated background there are muon induced activities that emits $\beta$-neutron cascade. Since all such cases have lifetimes $\tau \leq 1$ s they can be vetoed by the muon signal. A signal rate as low as $\sim$ 1 $\bar{\nu}_{e}$ 
event per year an 300 ton appears mesurable.

The signal energy is $E(MeV) = E(\bar{\nu}_{e}) - 1.8 + 2m_{e}c^{2} = E(\bar{\nu}_{e}) - 0.78$. Thus, even "at threshold" 
$\bar{\nu}_{e}$'s procude an $e^{+}e^{-}$ signal of 1.02 MeV.

The interesting sources of $\bar{\nu}_{e}$ are Supernovae, possible $\bar{\nu}_{e}$ emiting from the Sun, $\bar{\nu}_{e}$ emitted by the Earth and from nuclear power reactors.

The liquid scintillator is composed by pseudocumene (PC) as solvent and PPO at the concentration 
of 1.5 g/l as solute. The pseudocumene is 1,2,4-trimethylbenzene $C_6H_3(CH_3)_3$, and the PPO
is 2,5-diphenyloxazole $C_{15}H_{11}NO$.

\section{Supernovae neutrinos}

In a liquid scintillator detector, the electron antineutrino on proton reaction  $ \bar{\nu}_{e} + p \rightarrow n + e^+ $ constitute the majority of the detected Supernovae neutrino events. The presence on carbon in the organic scintillator 
provides an additional interesting target for neutrino interactions. Table \ref{tb:tableneutrinoson12C} summarizes the neutrino interactions on $^{12}C$ nucleus.

\begin{table}[htb]
\caption{
Neutrino interactions on $^{12}C$ nucleus \protect\cite{fukugita}.
}
\label{tb:tableneutrinoson12C}
\begin{center}
\begin{tabular}{ccc}\hline
& Reactions	& Threshold \\ \hline
$^{12}B_{gs}$ & $ \bar{\nu}_{e} + ^{12}C \rightarrow ^{12}B + e^+ $ & 14.4 MeV \\
$^{12}N_{gs}$ & $ \nu_{e} + ^{12}C \rightarrow ^{12}N + e^- $       & 17.3 MeV \\
$^{12}C^{*}$  & $ \nu + ^{12}C \rightarrow ^{12}C^{*} + \nu $       & 15.1 MeV \\
\hline
\end{tabular}
\end{center}
\end{table}

The neutrino interactions on $^{12}C$ can be tagged. The charged current events (CC) have a delayed coincidence of a $\beta$ 
decay following the interaction: $ ^{12}B \rightarrow ^{12}C + e^- +\bar{\nu}_{e}$ ($\tau_{1/2}$ = 20.2 ms) and
$ ^{12}N \rightarrow ^{12}C + e^+ +{\nu}_{e}$ ($\tau_{1/2}$ = 11.0 ms). The neutral current events (NC) have a monoenergetic $\gamma$-ray at 15.1 MeV: $ ^{12}C^* \rightarrow ^{12}C + {\gamma}$ (15.1 MeV).

\subsection{Supernovae neutrino bursts}
A Supernovae explosion releases $\sim$ $3 \cdot 10^{53}$ ergs of gravitational binding energy; the bulk of the energy is released 
in the form of neutrinos. A first $\nu_e$ burst is emitted during the infall stage of the collapse via the electron capture by proton: 
$e^- + p \rightarrow \nu_e +n$. After the core bounce, also neutrinos of other flavours are produced via nucleonic bremsstrahlung and 
pair annihilation processes as: $e^+ + e^- \rightarrow \nu + \bar{\nu}$.

The total emitted energy is equally shared by all six neutrino flavors:  $L_{\nu_{e}} \approx L_{\bar\nu_{e}} \approx L_{\nu_{x}}$ where $\nu_{x}$ stands for $\nu_{\mu}, \nu_{\tau}, \bar{\nu}_{\mu}, \bar{\nu}_{\tau}$.

\begin{figure}
\begin{minipage}[b]{8.5cm}
\centering
\includegraphics[width=6cm]{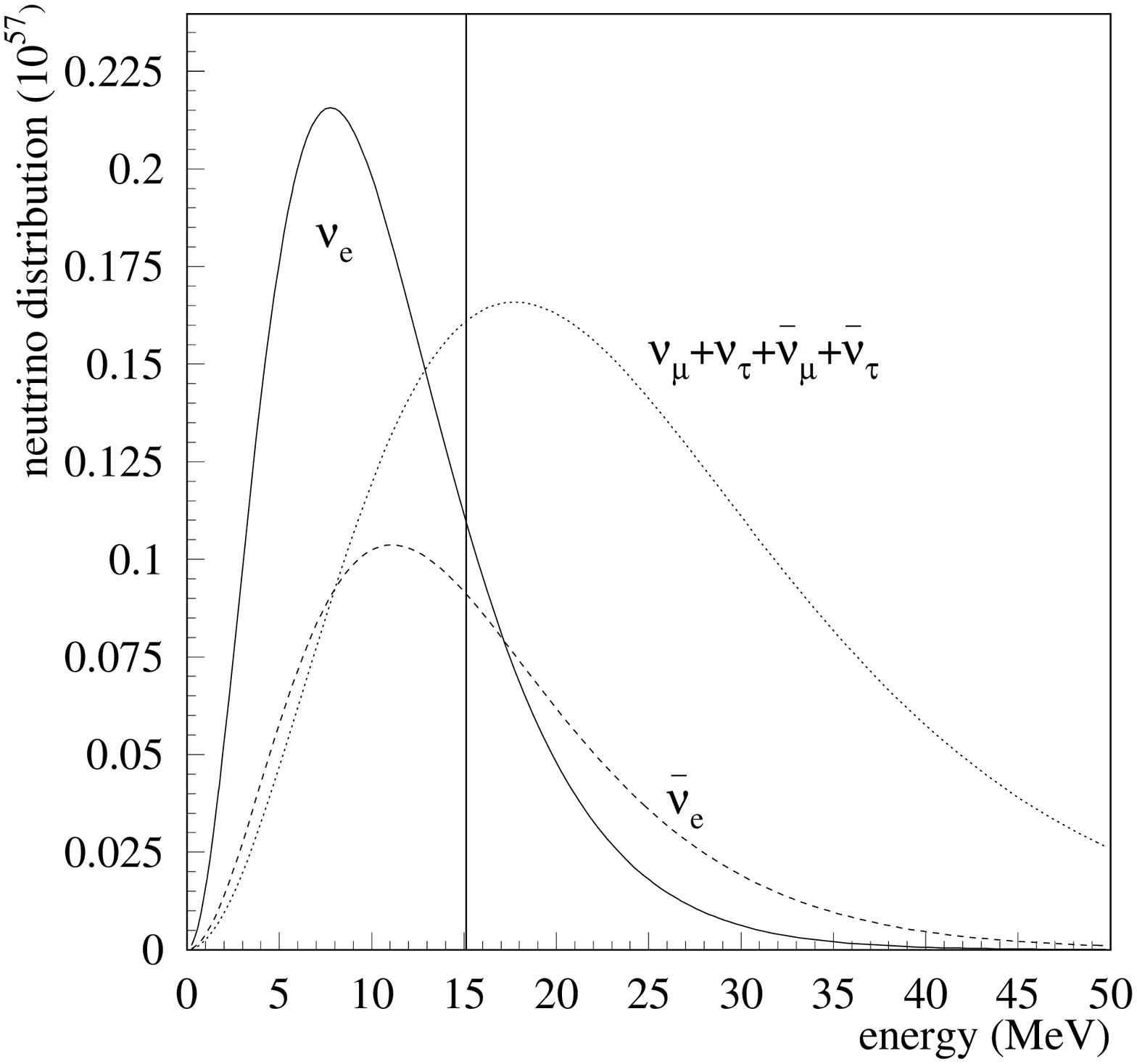}
\caption{Supernovae neutrino energy spectra: $\nu_{e}$,   $\bar{\nu}_{e}$,  ($\nu_{\mu}, \nu_{\tau}, \bar{\nu}_{\mu}, \bar{\nu}_{\tau}$).}
\label{fg:fig3a}
\end{minipage}
\ \hspace{2mm} \hspace{3mm} \
\begin{minipage}[b]{8.5cm}
\centering
\includegraphics[width=6cm]{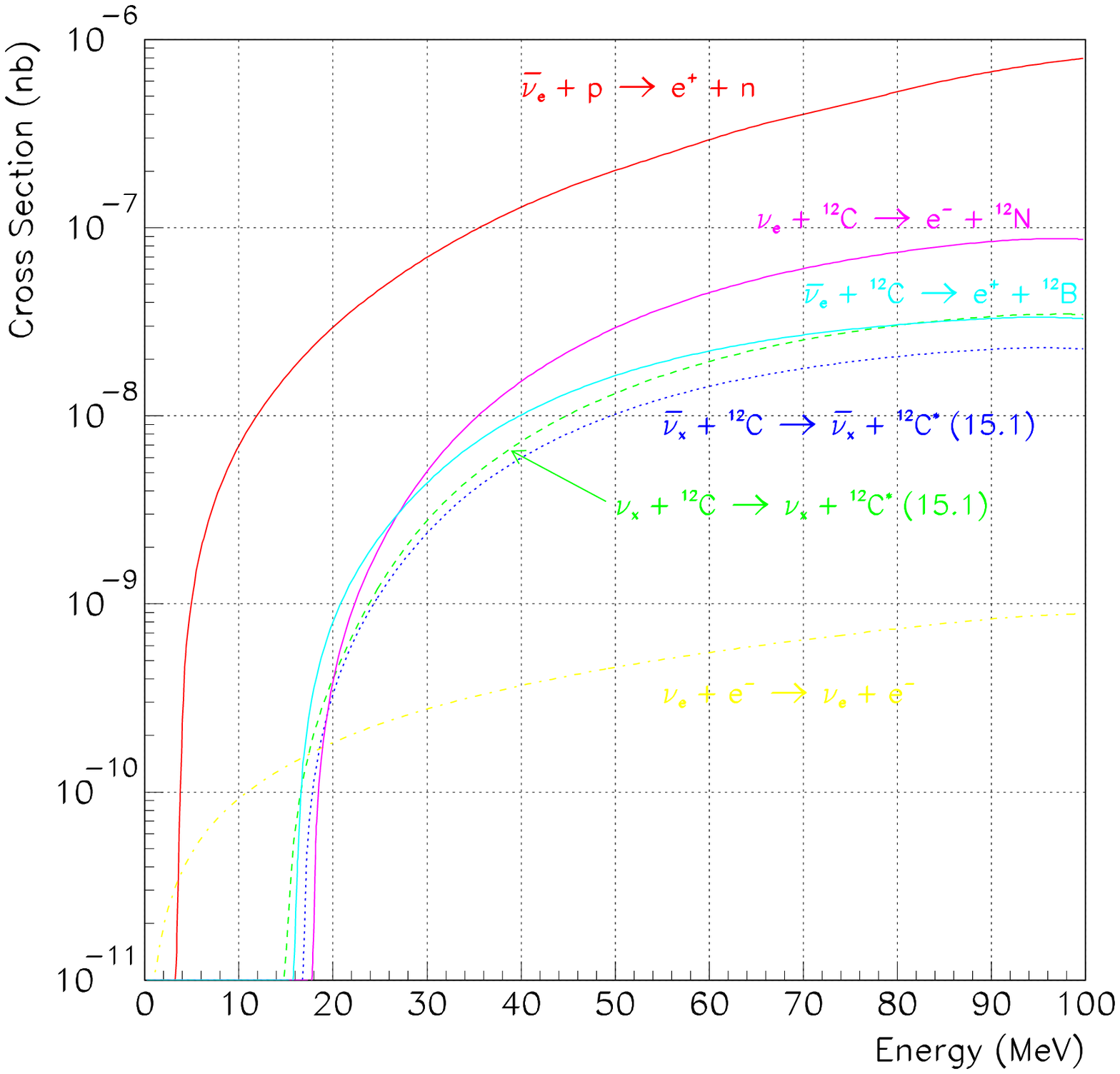}
\caption{Cross sections for CCp, ES, CC and NC on $^{12}C$.}
\label{fg:scint_cross}
\end{minipage}
\end{figure}

Since $\nu_{\mu}$ and $\nu_{\tau}$ interact in ordinary matter only via the neutral current weak interaction, they decouple at higher temperature. Furthermore, the neutrino decoupling takes place in neutron rich matter, which is less transparent to $\nu_e$ than $\bar{\nu}_{e}$. For this reason we will have a temperature hierarchy: $T_{\nu_{e}} \leq T_{\bar\nu_{e}} \leq T_{\nu_{x}}$; so, the neutrino mean energies are expected to be: $\langle E_{\nu_{e}} \rangle \simeq 11 MeV $, $\langle E_{\bar\nu_{e}} \rangle \simeq 16 MeV $, $\langle E_{\nu_{x}} \rangle \simeq 25 MeV $ \cite{langanke}. Figure \ref{fg:fig3a} shows the Supernovae neutrino energy spectra.

\subsection{Supernovae neutrino events in Borexino}

We consider 300 tons of pseudocumene and a type II Supernovae at 10 kpc from the Earth (that is roughly the center of the Galaxy).

Supernovae neutrino interact in the organic liquid scintillator via electron scattering, inverse $\beta$ decay of the proton (i.e. Cowan Reines reaction) and reactions on $^{12}C$ \cite{ArticoloCadonati}, \cite{Beacom05301} and \cite{Beacom093012}.

{\bf [ES]} The neutrino electron scattering gives recoil electrons with energy from zero to kinematic maximum. In Borexino the threshold is 250 keV due to the $\beta$ spectrum of the $^{14}C$ which is small compared to the Supernovae neutrino energies. This reaction is possible for all kind of neutrinos; $\nu_{e}$ and $\bar{\nu}_{e}$ interact with electrons via CC (with different couplings) and NC reactions, whereas $\nu_{x}$ interact only via NC interactions. The $\nu_{e}$ have the highest cross section:
$\sigma_{{{\nu}_{e}-e}} = 9.2 \cdot 10^{-45} T_{\nu_{e}} (MeV)$ $cm^2$ with this cross section hierarchy: 
$\sigma_{{{\nu}_{e}-e}} \approx 3\cdot\sigma_{{{\bar\nu}_{e}-e}} \approx 6\cdot\sigma_{{{\nu}_{x}-e}} $. The calculated events rate in Borexino, from neutrino electron scattering, is about five events (See Table \ref{tb:eventsfromES}).

\begin{table}[htb]
\caption{
Neutrino interactions from elastic scattering in Borexino \protect\cite{ArticoloCadonati}, \cite{Beacom05301} and \cite{Beacom093012}.
}
\label{tb:eventsfromES}
\begin{center}
\begin{tabular}{cc}\hline
Reactions	& $N_{events}$ \\ \hline
$ {{{\nu}_{e}-e}}     $ & 2.37 \\
$ {{{\bar\nu}_{e}-e}} $ & 0.97 \\
$ {{{\nu}_{x}-e}}     $ & 0.81 \\
$ {{{\bar\nu}_{x}-e}} $ & 0.67 \\
\hline
\end{tabular}
\end{center}
\end{table}

{\bf [Inverse $\beta$ decay]} The inverse $\beta$ decay ($ \bar{\nu}_{e} + p \rightarrow n + e^+ $) expected in Borexino will be about 80.

{\bf [Reactions on $^{12}C$]} For the neutrino reactions on $^{12}C$, cross section have been investigated theoretically and experimentally since 1980's. Measurements have been performed at KARMEN \cite{KARMEN}, LAMPF \cite{LAMPF} and LSND \cite{LSND}. Figure \ref{fg:scint_cross}  shows the cross section values for CCp, ES, CC and NC on carbon processes.

\begin{figure} 
\centering
\includegraphics [scale=0.30] {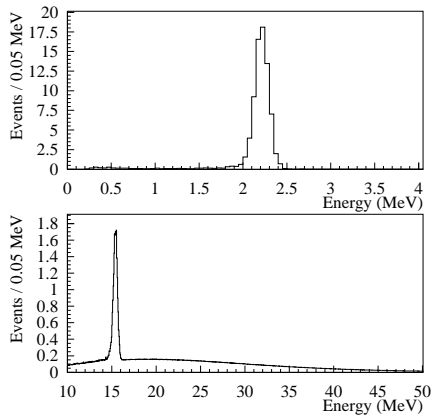}
\caption{Simulated spectrum from Supernovae neutrino in Borexino for two different energy regions.}
\label{fg:fig2}
\end{figure}

Table \ref{tb:eventsfromCCandNCon12C} summarizes the expected Supernovae neutrino events in Borexino for the CC and NC on $^{12}C$.
Since the $\nu_\mu$ and $\nu_\tau$ are more energetic than $\nu_e$ they dominate the neutral current reactions $ \nu + ^{12}C \rightarrow ^{12}C^{*} + \nu $ with an estimated contribution of about $90\%$. In order to exploit these aspects, a liquid scintillator Supernovae neutrino detector needs to be able to cleanly detect the 15.1 MeV $\gamma$-ray. Figure \ref{fg:fig2} shows the simulated spectrum from Supernovae neutrino in the Borexino detector.

\begin{table}[htb]
\caption{Neutrino interactions from elastic scattering in Borexino \protect\cite{ArticoloCadonati}, \cite{Beacom05301} and \cite{Beacom093012}.}
\label{tb:eventsfromCCandNCon12C}
\begin{center}
\begin{tabular}{cc}\hline
Reactions	& $N_{events}$ \\ \hline
$ \nu_{e} + ^{12}C \rightarrow ^{12}N + e^- $          &  0.7 \\
$ \bar{\nu}_{e} + ^{12}C \rightarrow ^{12}B + e^+ $    &  3.8 \\
${\nu}_{e} + ^{12}C$                                   &  0.4 \\
$\bar{\nu}_{e} + ^{12}C$                               &  1.5 \\
${\nu}_{x} + ^{12}C$                                   & 20.6 \\
\hline
\end{tabular}
\end{center}
\end{table}

By studying the arrival time of neutrinos of different flavors from a Supernovae, mass limit on $\nu_\mu$ and $\nu_\tau$ down to some 10 of eV level can be explored. The time delay, in Borexino, is obtained by measuring the time delay between NC events and CC events \cite{ArticoloCadonati}, \cite{Beacom05301} and \cite{Beacom093012}. 
In a detector with sufficiently low threshold, as in the case of  Borexino, the elastic scattering on proton can be observed by detecting the recoiling protons; this reaction could furnish spectroscopic information among the neutral reactions \cite{Beacom033001}.

\section{Neutrinos from the Earth}

Our planet emits a very small flux of heat with an average enegy value of $\Phi_H \sim 80$ $mW/m^2$. Intergrating over the whole Earth surface we obtained a total heat $H_E \sim 40$ $TW$. 

The source of this energy is not understood qauntitatively. It is possible to study the radiochemical composition of the Earth by detecting antineutrino ($\bar{\nu}_{e}$) emitted by the decay of radioactive isotopes; confirming the abundance of certain radioelements could establish important geophysical constraints on the heat generation within the Earth.
The main contribution of the radiogenic heat is due to the two decay chains $^{238}U$ and $^{232}Th$ and the decay of the $^{40}K$ (we neglet $^{235}U$ and $^{87}Rb$ which provide smaller contributions). Thanks to the measurement of the antineutrino flux it would be possible to provide a direct information about the composition and the amount of the radioactive material within the Earth leading to the determination of the radiogenic contribution to the heat flow.

The energy threshold of the reaction $\bar{\nu}_{e} + p \rightarrow n + e^+$ is 1.8 MeV. There are 4 $\beta$-decay in the $^{238}U$ and $^{232}Th$ chains with energy $\geq$ 1.8 MeV: $^{214}Bi$ [U] $\leq 3.27$ MeV, $^{234}Pa$ [U] $\leq 2.29$ MeV, $^{228}Ac$ [Th] $\leq 2.08$ MeV and $^{212}Bi$ [Th] $\leq 2.25$ MeV.

The signal energy is $E(MeV) = E(\bar{\nu}_{e}) - 1.8 + 2m_ec^2$. The terrestrial antineutrino spectrum above 1.8 MeV has a "2-component" shape: the high energy component coming solely from U chain and the low energy component coming with contributions from U and Th chains. This signature allows individual assay of U and Th abundance in the Earth.

\subsection{Equation for heat and neutrino luminosity}

In Table \ref{tb:radioelements} are presented the heat contribution and the neutrino luminosity from the three natural radioelements. We remind that the number of disintegration per second (i.e. Bq) per gram for the three elements are: $\sim$ 12300 Bq/g for $^{238}U$, $\sim$ 4020 Bq/g for $^{238}Th$ and $\sim$ 29.8 Bq/g for $^{40}K$ ($^{40}K$ = 0.0118$\%$ of $^{nat}K$).

\begin{table}[htb]
\caption{
Power and neutrino luminosity from the three long lived radioelemets ($\epsilon$ is the present natural isotopic abundance).
}
\label{tb:radioelements}
\begin{center}
\begin{tabular}{ccc}\hline
Reactions & [$W \cdot g^{-1}$]	& [$\bar{\nu}_{e} \cdot g^{-1} \cdot s^{-1}$] \\ \hline
$^{238}U$ $\longrightarrow$ $^{206}Pb$ + $8\alpha$ + $6e^-$ + $6\bar{\nu}_{e}$ + 51.7 MeV
 & $\epsilon(U)$ $\simeq$ $9.5 \cdot 10^{-8}$ 
 & $\epsilon_{\bar{\nu}_{e}} (U)$ $\simeq$ $7.4 \cdot 10^4$  \\
$^{232}Th$ $\longrightarrow$ $^{208}Pb$ + $4\alpha$ + $4e^-$ + $4\bar{\nu}_{e}$ + 42.8 MeV
 & $\epsilon(Th)$ $\simeq$ $2.7 \cdot 10^{-8}$ 
 & $\epsilon_{\bar{\nu}_{e}} (Th)$ $\simeq$ $1.6 \cdot 10^4$  \\
$^{40}K$ $\longrightarrow$ $^{40}Ca$ + $e^-$ + $\bar{\nu}_{e}$ + 1.32 MeV (89 $\%$)
 & $\epsilon(K)$ $\simeq$ $3.6 \cdot 10^{-12}$       
 & $\epsilon_{\bar{\nu}_{e}} (K)$ $\simeq$ $27$  \\
$^{40}K$ + $e^-$ $\longrightarrow$ $^{40}Ar$ + $\nu_{e}$ + 1.51 MeV (11 $\%$)
 & $\epsilon(K)$ $\simeq$ $3.6 \cdot 10^{-12}$        
 & $\epsilon_{\nu_{e}} (K)$ $\simeq$ $3.3$  \\
\hline
\end{tabular}
\end{center}
\end{table}

The heat production rate (H) and the neutrino luminosity (L) are given by \cite{fiorentini}: $H [W] = 9.5 \cdot 10^{-8} M(U) + 2.7 \cdot 10^{-8} M(Th) + 3.6 \cdot 10^{-12} M(K)$; $L_{\bar{\nu}_{e}} [\bar{\nu}_{e}/s] = 7.4 \cdot 10^{4} M(U) + 1.6 \cdot 10^{4} M(Th) + 27 M(K)$; $L_{{\nu}_{e}} [{\nu}_{e}/s] =  3.3 M(K)$ where M(U), M(Th) and M(K) are the masses of the elements in grams. Everything is fixed in term of 3 numbers: M(U), M(Th) and M(K) or M(U), Th/U and K/U. With these simple equations it is possible to determine the radiogenic heat production and the neutrino flow from model of the Earth composition.

\subsection{Terrestrial ${\bar{\nu}_{e}}$ events in Borexino}

The starting point for determining the distribution of U, Th and K in the present crust and mantle is understanding the composition of the "Bulk Silicate Earth" or BSE for short, which is the model representing the primordial mantle prior to crust formation consistent with observation and geochemistry (equivalent in composition to the modern mantle plus crust).
BSE concentrations of 20 ppb ($\pm 20 \%$), $Th/U \sim 3.8$ and $K/U \sim 10^4$ have been suggested \cite{kargel}.
Since the mantle mass is $68 \%$ of the mass Earth ($M_{Earth} = 6 \cdot 10^{27} g$), in the BSE model we obtain a radiogenic heat production rate of about 20 TW ($\sim 8$ TW from U, $\sim 8.6$ TW from Th and $\sim 3$ TW from K); the antineutrino production is dominated by K.

\begin{figure}
\begin{minipage}[b]{8.5cm}
\centering

\includegraphics [scale=0.3] {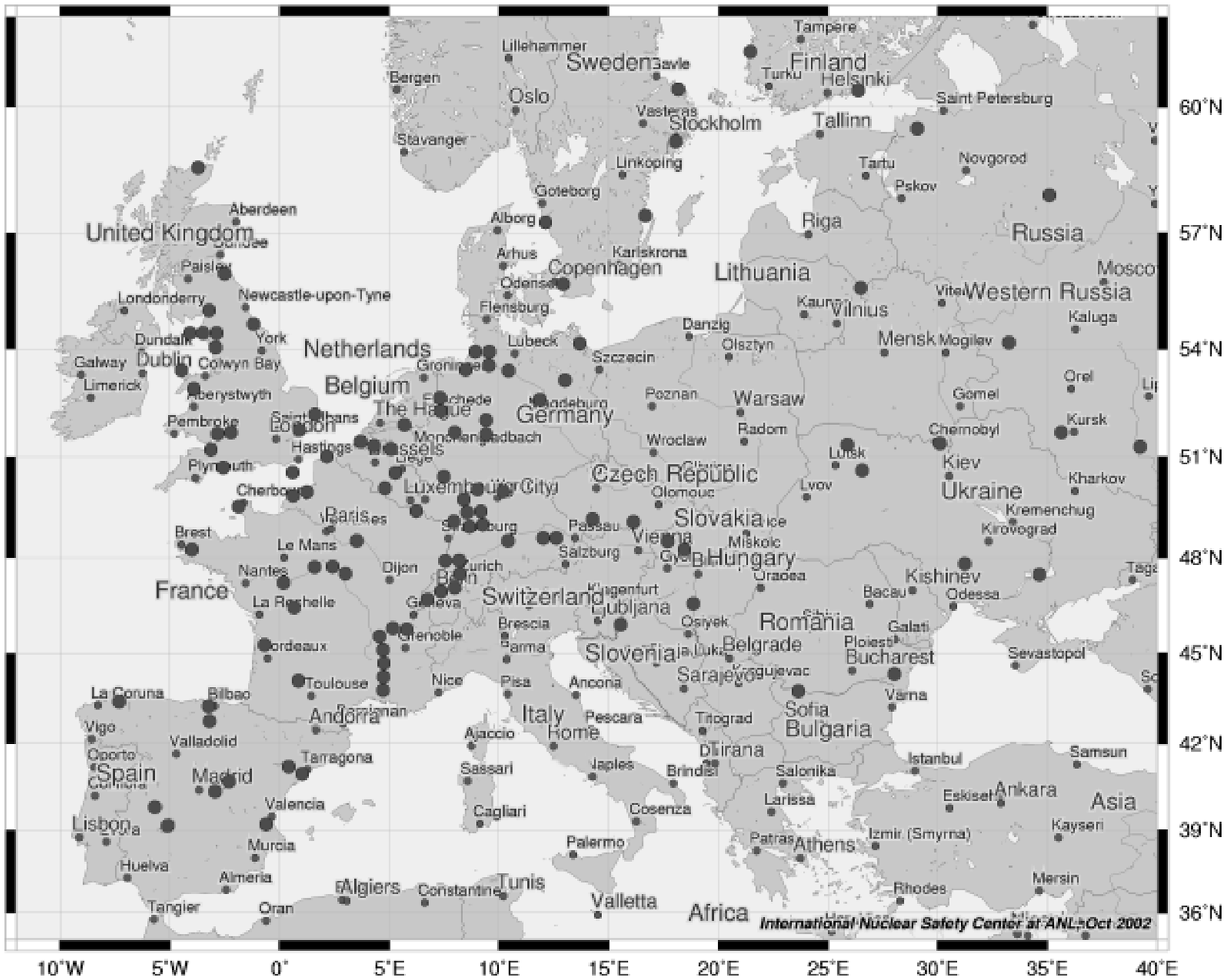}
\caption{Maps of Nuclear Power Reactors in Europe \cite{INSC}.}
\label{fg:europereactors}

\end{minipage}
\ \hspace{2mm} \hspace{3mm} \
\begin{minipage}[b]{8.5cm}
\centering

\includegraphics [scale=0.4] {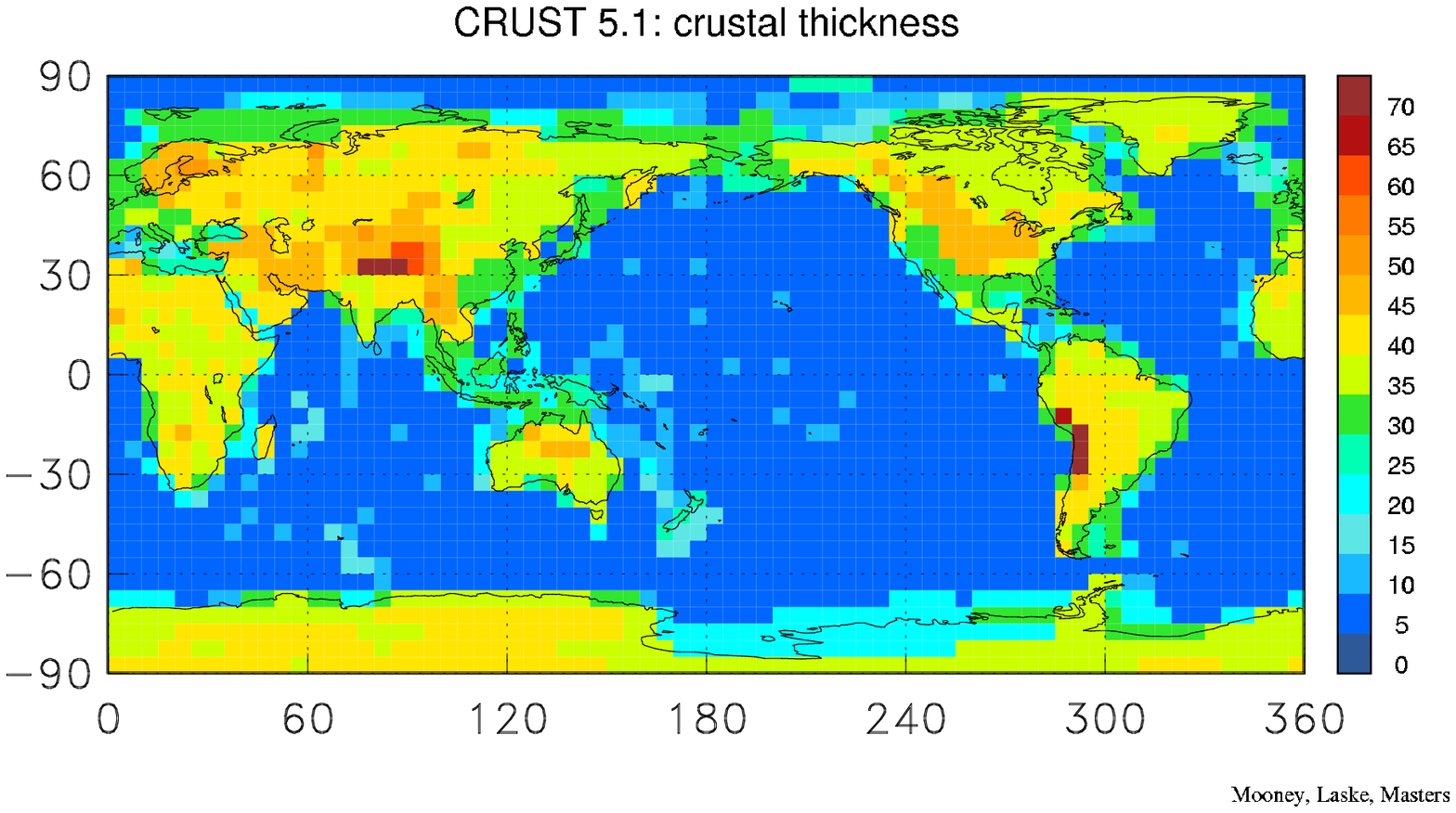}
\caption{Crustal thickness from the global crustal model \cite{GCM}.}
\label{fg:thick2}

\end{minipage}
\end{figure}

During the formation of the Earth's crust, the primitive mantle was depleted in U, Th and K, while the crust\footnote{The Continental Crust(CC) has an average thickness of $\sim 40$ km while Oceanic Crust (OC) has an average thickness of $\sim 6$ km. Furthermore CC is about 10 times richer in U and Th than OC.} was enriched. Measurements of the crust provide isotopic abundance information. With these measurement, it is possible to deduce the average U and Th concentrations in the present depleted mantle. A crust type and tickness data in the form of a global crustal map has been employed (see Figure \ref{fg:thick2}).

A background that cannot be shielded is the $\bar{\nu}_{e}$ from nuclear plants. In Italy there are not active nuclear power installations, thus the background induced is not so high. In order to estimate this background contribution it has been employed data from the International Nuclear Safety Center which provide positions and power of all reactors operating in the world (see Figure \ref{fg:europereactors}).

\begin{figure} 
\centering
\includegraphics [scale=0.4] {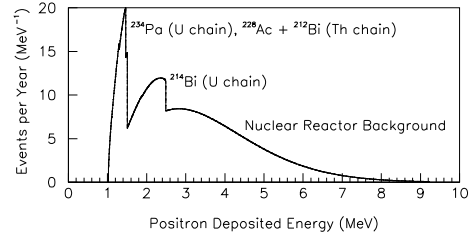}
\caption{Positron energy spectrum from ${\bar{\nu}_{e}}$ events in the Borexino detector \cite{calaprice}.}
\label{fg:geoneutrinos}
\end{figure}

Borexino is homed in the Gran Sasso underground laboratory (LNGS) in the center of Italy: 42$^{\circ}$N 14$^{\circ}$E. The calculated ${\bar{\nu}_{e}}$ flux at the Gran Sasso laboratory is $5.9 \cdot 10^{6}$ $cm^2 s^{-1}$, while the background flux is $0.65 \cdot 10^{6}$ $cm^2 s^{-1}$. Figure \ref{fg:geoneutrinos} shows the positron energy spectrum from antineutrino events in Borexino. The expected signal events are $\sim 7.8$ per year against a background of $\sim 29$ per year due to ${\bar{\nu}_{e}}$ from reactors, 7.6 of them in the same spectral region as the terrestrial ${\bar{\nu}_{e}}$.

The characteristic 2-component shape of the terrestrial anti-neutrino energy spectrum makes it possible to identify these events above the reactor anti-neutrino background. Since the reactor anti-neutrino background has a well-known shape it can be easily subtracted allowing the discrimination of the U contribution from the Th contribution.

\section{Conclusions}

The main characteristics that make Borexino interesting (for neutrino physics other than solar neutrinos) are for what concern Supernovae, the very effective ability to detect the high energy gamma peak (15.1 MeV) from NC reactions on $^{12}C$, this is because the unsegmented large volume detector. For what concern geoneutrinos, the very low contribution to the antineutrino background due to the absence of nuclear plants in Italy.

\section{Acknowledgements}

I would like to thank professor Hiro Ejiri for his kind hospitality at the "The 1st Yamada Symposium
on Neutrinos and Dark Matter in Nuclear Physics". This work has been supported by the Italian National Institute of Nuclear Physics (INFN).


\begin{thebibliography}{99}

\bibitem{ArticoloCadonati} L.Cadonati, F.P.Calaprice, M.C.Chen,
Supernovae neutrino detection in Borexino
Astroparticle Physics 16 (2002) 361-372.

\bibitem{Beacom05301} J.F.Beacom, P.Vogel,
Phys. Rev. D58 (1998) 053010.

\bibitem{Beacom093012} J.F.Beacom, P.Vogel,
Phys. Rev. D58 (1998) 093012.


\bibitem{Beacom033001} J.F.Beacom, W.Farr, P.Vogel,
Phys. Rev. D66 (2002) 033001.



\bibitem{calaprice} C.G.Rothschild, F.P.Calaprice, M.C.Chen,
Antineutrino geophysics with liquid scintillator detector
arXvi:nucl-ex/9710001 v1 15 Oct 1997.


\bibitem{STBX} Borexino Collaboration,
Astroparticle Physics 16 (2002) 205-234.


\bibitem{fukugita} M.Fukugita, Y.Kohyama, K. Kubodera,
Phys. Lett. B 212 (1988) 139.

\bibitem{langanke} K.Langanke, P.Vogel, E.Kolbe,
Phys. Rev. Lett. 76 (1996) 2629.

\bibitem{KARMEN} KARMEN Collaboration,
Prog. Part. Nucl. Phys. 32 (1994) 351.

\bibitem{LAMPF} D.A.Krakauer et al.,
Phys. Rev. C 45 (1992) 2450.

\bibitem{LSND} C.Athanassopoulos et al.,
Phys. Rev. C 55 (1997) 2078.

\bibitem{fiorentini} Neutrinos and Energetics of the Earth,
arXvi:nucl-ex/0212008 v2 13 Feb 2003.

\bibitem{kargel} J.S.Kargel and J.S.Lewis,
The composition and early evolution of Earth, Icarus 105, 1-25, 1993.

\bibitem{GCM} W.Mooney G.Laske and T.G.Masters,
http://quake.wr.usgs.gov/study/CrustalStructure/

\bibitem{INSC} International Nuclear Safety Center,
http://www.insc.anl.gov



\end{thebibliography}
\end{document}